\documentclass[10pt,conference]{IEEEtran} 

\usepackage{graphicx} 

\usepackage[utf8]{inputenc}
\usepackage[T1]{fontenc}
\usepackage{balance}
\usepackage{microtype}
\usepackage{fix-cm}
\usepackage{paralist}
\usepackage{balance}
\usepackage{acronym}

\usepackage[UKenglish]{babel}
\acrodef{AODV}{Ad hoc On-demand Distance Vector}
\acrodef{AP}{Access Point}
\acrodef{API}{Application Programming Interface}
\acrodef{BATMAN}{Better Approach To Mobile Adhoc Networking}
\acrodef{BSS}{Basic Service Set}
\acrodef{BT}{Bluetooth}
\acrodef{CCN}{Content Centric Network}
\acrodef{DNA}{Distributed Numbering Architecture}
\acrodef{DHCP}{Dynamic Host Configuration Protocol}
\acrodef{DS}{Distribution System}
\acrodef{GO}{Group Owner}
\acrodef{GPS}{Global Positioning System}
\acrodef{HWMP}{Hybrid Wireless Mesh Protocol}
\acrodef{IBSS}{Independent Basic Service Set}
\acrodef{IEEE}{Institute of Electrical and Electronics Engineers}
\acrodef{IETF}{Internet Engineering Task Force}
\acrodef{IPC}{Inter-Process Communication}
\acrodef{MAC}{Medium Access Control}
\acrodef{MANET}{Mobile AdHoc NETwork}
\acrodef{MBSS}{Mesh Basic Service Set}
\acrodef{MDP}{Mesh Datagram Protocol}
\acrodef{OLSR}{Optimized Link State Routing}
\acrodef{OS}{Operating System}
\acrodefplural{OS}[OS's]{Operating Systems}
\acrodef{PLC}{Power Line Communication}
\acrodef{QoS}{Quality of Service}
\acrodef{RTT}{Round-Trip Time}
\acrodef{SAE}{Simultaneous Authentication of Equals}
\acrodef{SSID}{Service Set Identifier}
\acrodef{UDP}{User Datagram Protocol}
\acrodef{VoMP}{Voice Over Mesh Protocol}
\acrodef{VPN}{Virtual Private Network}
\acrodef{WLAN}{Wireless Local Area Network}
\acrodef{WMN}{Wireless Mesh Network}
\acrodef{DNS}{Domain Name System}

\newcommand\figref[1]{Fig.~\ref{fig:#1}}
\newcommand\tabref[1]{table~\ref{tab:#1}}

\usepackage[pdfpagelabels=false, hidelinks, bookmarks=false,
pdftitle={Experimentation with MANETs of Smartphones},
pdfkeywords ={Ad-hoc networks, MANET, mesh networks, Android},
pdfsubject = {},
bookmarksnumbered=false,
pdfauthor ={Eduardo Soares, Pedro Brandao, Rui Prior, Ana Aguiar },
breaklinks=true,
draft
]
{hyperref} 
\usepackage{needspace}

\usepackage{soul}

\usepackage[usenames,dvipsnames]{color}
\definecolor{gray-cell}{gray}{.80}
\definecolor{LightRed}{rgb}{1,0.78,0.58}
\definecolor{LRed}{rgb}{1,0.5,0}
\definecolor{gray-pm}{gray}{.50}
\definecolor{gray-xm}{gray}{.60}
\usepackage{colortbl}

\usepackage{listings}

\usepackage{amsfonts}
\usepackage{threeparttable}
\usepackage{pifont}
\newcommand{\cmark}{\ding{51}}%
\newcommand{\xmark}{\ding{53}}%
\newcommand{\pmark}{\ding{117}}%
\newcommand{\cmarkcell}{\cmark}%
\newcommand{\xmarkcell}{\cellcolor{LRed}\xmark}%
\newcommand{\pmarkcell}{\cellcolor{LightRed}\pmark}%

\author{
	\IEEEauthorblockN{Eduardo~Soares\IEEEauthorrefmark{1}, Pedro~Brandão\IEEEauthorrefmark{1}, Rui~Prior\IEEEauthorrefmark{1}, Ana~Aguiar\IEEEauthorrefmark{2}}
  \IEEEauthorblockA{%
    \IEEEauthorrefmark{1}Instituto de Telecomunicações and Faculdade de Ciências da Universidade do Porto \\\{esoares, pbrandao, rprior\}@dcc.fc.up.pt\\
    \IEEEauthorrefmark{2}Instituto de Telecomunicações and Faculdade de Engenharia da Universidade do Porto \\anaa@fe.up.pt
  }
}

\begin{document}

\title{Experimentation with MANETs of Smartphones}

\maketitle

\begin{abstract}
\acp{MANET} have been identified as a key emerging technology for scenarios in which IEEE 802.11 or cellular communications are either infeasible, inefficient, or cost-ineffective. Smartphones are the most adequate network nodes in many of these scenarios, but it is not straightforward to build a network with them. 
We extensively survey existing possibilities to build applications on top of ad-hoc smartphone networks for experimentation purposes, and introduce a taxonomy to classify them.
We present AdHocDroid, an Android package that creates an IP-level \ac{MANET} of (rooted) Android smartphones, and make it publicly available to the community. AdHocDroid supports standard TCP/IP applications, providing real smartphone IEEE 802.11 \ac{MANET} and the capability to easily change the routing protocol. We tested our framework on several smartphones and a laptop. We validate the MANET running off-the-shelf applications, and reporting on experimental performance evaluation, including network metrics and battery discharge rate.
\end{abstract}




\section{Introduction}

Although Internet connectivity is nearly ubiquitous, there are many situations in which using infrastructureless communication is better than an IEEE 802.11 hotspot or cellular communication, because the latter are either infeasible, inefficient, or cost ineffective. For example, in remote areas, e.g. forests, ocean, or in catastrophe scenarios~\cite{Mitra2012ICC,Srikrishna2012}, there is simply no infrastructure to provide connectivity. Or in social upraise scenarios, in which the infrastructure cannot be trusted, as the use of Open Garden\footnote{http://opengarden.com/about} 
shows. A more leisurely example is low latency gaming~\cite{Huang2011,Le2012} or sharing a file with acquaintances. Another application scenario could be group communication in mass events, like conferences or concerts~\cite{Turkes2015}, in which infrastructure may be unable to support all communication demand. For all these reasons, wireless ad-hoc networking was identified as a major emerging technology at the "Internet on the Move" workshop~\cite{Sathiaseelan2012a} and by Conti et al.~\cite{Conti2014}. 

Although smartphones are privileged network nodes in the people centric scenarios mentioned above, work in this direction lies primarily in the field of middleware for  distributed applications, whereby connectivity is blindly assumed up to few exceptions like Haggle~\cite{Su2007Haggle}, and only a few provide support for smartphones. 
Recently, application frameworks that leverage WiFi Direct or cooperation between IEEE 802.11 and \ac{BT} appeared. 
But those solutions do not create an IP-level network, require overlay routing for multi-hop communication, and applications need to be adapted to a specific framework \acp{API}. 
IP-level multi-hop networking makes the difference in scenarios where latency is critical, and where communication with other IP enabled devices like laptops or PCs is wanted. Moreover, it is completely transparent to applications, which just use the sockets \ac{API}.

In this paper, we review work on \acp{MANET} of smartphones, and proceed to extensively survey solutions that claim to provide ad-hoc connectivity for smartphones (section~\ref{sec:related-work})
We then introduce AdHocDroid to turn smartphones into nodes of an IP-level mobile ad-hoc network. A \ac{MANET} of Android smartphones enables simplified instantiation of a test-bed for experimental evaluation of routing protocols, data dissemination, distributed applications, etc.~\cite{Kiess2007}. 
We describe how to set up an IP-level 802.11 ad-hoc network with multi-hop capability on smartphones running the Android \ac{OS}, and share lessons learned (section~\ref{sec:adhocdroid}). Our purpose is to advance MANET experimentation by enabling easy testing on MANET network protocols.
Then, we introduce a taxonomy of \ac{MANET} features and use it to characterize the surveyed solutions and AdHocDroid (section~\ref{sec:manetcompare}). 
Finally, we experimentally validate AdHocDroid running applications on the network, and evaluating network performance (throughput and latency) and battery consumption (section~\ref{sec:performance-evaluation}).

\section{Related Work}
\label{sec:related-work}
A \ac{MANET} is a wireless ad-hoc network that allows, and adapts to, mobility of the participating nodes, which are terminals that also route packets of flows in which they are not endpoints. An ad-hoc network is any type of self-configuring network that does not require pre-existing installed and configured infrastructure. As such, nodes in a \ac{MANET} are able to communicate with every node in the \ac{MANET} as long as a path between the nodes may be established. 



\textit{SocialMesh} explores the idea of an ad-hoc network of citizen's devices as an infrastructure-less communication network, and its vulnerabilities to a wide range of attacks~\cite{Srikrishna2012}. Although smartphones are identified as straightforward instantiations of nodes, the article focuses on the security aspects and evaluates them qualitatively. Fully distributed services on top of ad-hoc networks have been explored in a plethora of works, focusing on middleware that provides service discovery~\cite{Mitra2012ICC}, multicast group communication~\cite{Mitra2012PMC,Huang2011}
, data dissemination~\cite{Turkes2015}, improved latency~\cite{Le2012}, or all of them and more, like Haggle~\cite{Su2007Haggle}. These works assume that multi-hop connectivity is provided either through IP or as overlay on IEEE 802.11 and\slash or \ac{BT}. In this sense, this paper addresses alternatives to provide the necessary underlying multi-hop connectivity at the network layer using smartphones.


A survey of experimental work with \acp{MANET}~\cite{Kiess2007} shows that real-world experiments can bring to light significantly different behaviours than \acp{MANET} simulation or even emulation. 
The survey describes 5 static experiments and 8 with mobility, but only 2 testbeds (APE and ORBIT), which use laptops running the Linux \ac{OS} with 802.11 dongles. The survey highlights the importance of real world-experimentation and describes toolsets. 
Recently, Papadopoulos et al.~\cite{papadopoulos2016} highlighted the benefits of experimentation for the deployment of ad-hoc networks, and identify reproducibility as a caveat of the methodology. In this sense, we expect to contribute to the acceleration of experimentation with ad-hoc networks of smartphones by providing a tool that simplifies setting up such a network.


The next sub-sections describe protocols and frameworks that aim at providing \acp{MANET}.

\subsection{802.11 Support}

The latest revision of the IEEE 802.11 standard~\cite{IEEE802112012} supports two different modes that can be used for ad-hoc networking: \ac{IBSS} mode and 802.11s.

\subsubsection{\ac{IBSS} Mode} 

The \ac{IBSS} mode of 802.11, commonly referred to as \emph{ad-hoc mode} because it does not require any infrastructure to be in place, can be used as a basis for mesh networking.  In this mode, all nodes play similar roles, and any node can communicate directly with any other node within the network, defined as the set of nodes in \ac{IBSS} mode sharing the same \ac{SSID} that are within its radio range.

The \ac{IBSS} mode itself, however, does not provide multi-hop capabilities.  There is no provision for path discovery and selection, nor for relaying frames to nodes out of the radio range of the original sender. In \ac{IBSS}-based ad-hoc networks, these functions must be performed by an additional protocol, like \acsu{OLSR}~\cite{RFC3626} or \acsu{BATMAN}
\footnote{https://www.open-mesh.org/projects/open-mesh/wiki}, usually at the network layer. 
Since connectivity is provided at the link layer (optionally with additional network layer support for multi-hop), IP-based applications work without any modification in an \ac{IBSS}-based network. Interoperation with non-Android systems works out-of-the-box for the single-hop case, and requires a routing protocol for multi-hop.

\subsubsection{802.11s} 

More recently, mesh networking support has been introduced in 802.11 through the 802.11s amendment, now incorporated in the standard~\cite{IEEE802112012}. 802.11s defines the \ac{MBSS} that provides a wireless \ac{DS}, either independent or extending the wired \ac{DS}, based on meshing at the link layer.  
802.11s defines a standard path metric and path selection protocol, \ac{HWMP}, though others can be used as long as all stations in the mesh agree.  \ac{HWMP} combines reactive routing derived from \ac{AODV}~\cite{RFC3561}
with root-based proactive routing for communication with the outside. 

In Linux, 802.11s is supported through the open802.11s implementation
\footnote{http://open80211s.org/open80211s/}. Open802.11s supports wireless chipsets with a driver using the software implementation of the MAC layer provided by the \texttt{mac80211} kernel module. Currently, most wireless chipset drivers use a hardware implementation of the MAC layer, and are, thus, unsupported.  The driver must also be mesh-enabled. 
802.11s is now merged in the mainstream kernel, but support for it must be configured, and an updated version of the \texttt{iw} tool is also required. 

In Android, vendor-provided system images do not usually support 802.11s in the kernel or the configuration tools. This means that the use of 802.11s, even on devices with supported chipsets, is limited to those using third party, customized Android versions, and we are not aware of any that currently supports 802.11s.

\subsection{WiFi Direct}

WiFi Direct is a technical specification~\cite{WiFiP2PSpecV15} 
of the WiFi Alliance that leverages existing standards to provide a convenient way for securely connecting devices without installed infrastructure, enhanced with features like peer and service discovery.  It is based on the \emph{infrastructure \acsu{BSS}} mode of 802.11.  One of the devices, selected through negotiation, will become the group owner (\ac{GO}) and act as an \ac{AP}.  This has the advantage of allowing legacy clients to connect to the \ac{GO}.  The \ac{GO} incorporates a \ac{DHCP} server for providing IP addresses to the Client nodes.

WiFi Direct imposes a star topology, with the \ac{GO} at the center.  While the specification mentions Concurrent Devices that can simultaneously connect to the infrastructure or be part of a different group (requiring these devices to support multiple MAC entities), additional protocols are required for routing. 
A significant disadvantage of WiFi Direct is that if the \ac{GO} leaves, the group is torn down and a new group must be established from scratch.  While these limitations are irrelevant in simple situations like a printer letting computers and other devices connect, they make WiFi Direct unsuitable as a basis for multi-hop networking.

To the best of our knowledge, the experiments to provide multi-hop support using WiFi Direct are limited to \ac{CCN} approaches where the routing is hidden using the search for the content.  Jung et al.~\cite{wifidirectCCN14} try to use the concurrent operation mode to have the card connected to more than one point, but this work is only tested through NS-3 simulations. 
Content-centric device-to-device routing has been tried in non-rooted Android smartphones using WiFi Direct~\cite{arxiv:CasettiCPVDG14}. 
However, WiFi Direct in Android assigns the same IP address (192.168.49.1) to the \ac{GO} of all groups. The authors proxy the connection through several nodes to circumvent this problem.  
These issues reveal the inadequacy of using WiFi Direct for multi-hop networks, particularly in Android.

\subsection{Open Garden}
\label{sub:opengarden}
Open Garden
is a software for Internet connection sharing on mobile devices using a mesh of \ac{BT} or WiFi Direct links.  It also allows communication between devices across multiple hops as long as the application uses OpenGarden's proprietary forwarding software. The FireChat application, from the same company, runs on top of Open Garden enabling a multi-hop messaging framework. From the scarce documentation and our tests using the software, we concluded that no IP-level connectivity that might be used by other applications is provided. 

Open Garden works by creating a \ac{VPN} to a \ac{BT} paired device also running the application. The other device terminates the \ac{VPN} tunnel and either forwards the request to another node or redirects the message to the local application that registered for it (most commonly FireChat)\footnote{We were unable to verify the level of node identification used. From our experiments, it seemed that this was carried in a proprietary message that the terminating point (Open Garden software) would interpret. One indication of this is that the IP address of the \ac{VPN} tunnels were the same on all devices.}. With this architecture, a multi-hop overlay network is established using \ac{BT} connections.

When a device has Internet connection (through cellular or 802.11 infrastructure) it can forward the requests received.  This, again, after the Open Garden software interprets and re-routes the data packets. 
Thus, Open Garden is also not an alternative to set up a \ac{MANET} test-bed of smartphones.

\subsection{Serval Project}
\label{sec:serval}
The Serval project \cite{Gardner-Stephen2011} provides a free and open-source software to allow mobile phones to communicate in the absence of phone towers and other infrastructure, targetting disaster situations and remote communities. The Serval Mesh application
\footnote{http://developer.servalproject.org/dokuwiki/} provides voice calls, text messaging and file sharing directly over IEEE 802.11 links between mobile devices.  It can be used for peer-to-peer communication through an 802.11 \ac{AP} or in an ad-hoc multi-hop topology without infrastructure support. The \ac{MANET} is implemented using an ad-hoc routing protocol over 802.11 in \ac{IBSS} mode. The project initially used \ac{BATMAN}
, but moved to an \textit{in-house} routing protocol. 

The project developed \ac{MDP}, a hybrid of network and transport layer protocol that shares some properties with \ac{UDP}, but with per-hop retransmission of packets for mitigating the cumulative end-to-end packet loss effect that can significantly affect the performance of multi-hop wireless environments.  \ac{MDP} can work over IP, or directly over link layer technologies\footnote{The intention of supporting \ac{MDP} over \ac{BT} is stated.}.  On top of \ac{MDP}, the project provides Rhizome, a resilient file distribution protocol that is used to transparently transport data across the mesh nodes.  It is used for transmitting messages or support other services, such as their \ac{VoMP}.  The project also defines a \ac{DNA} to identify and address the nodes with cryptographic IDs on the network.

The current application on the Google Play Store includes the Serval Mesh that provides the above functionality including the project's routing protocol. 
Currently the development is being driven for mobile phones and Android is the one currently supported with applications. 
The specificities of the protocols outlined above make the Serval approach unusable by applications that are not aware of their \ac{API} and sub-system.  This provides little to no flexibility as a \ac{MANET} test-bed. 

\section{AdHocDroid}\label{sec:adhocdroid}
\begin{figure}[!ht]
	\centering
    \includegraphics[width=.98\linewidth]{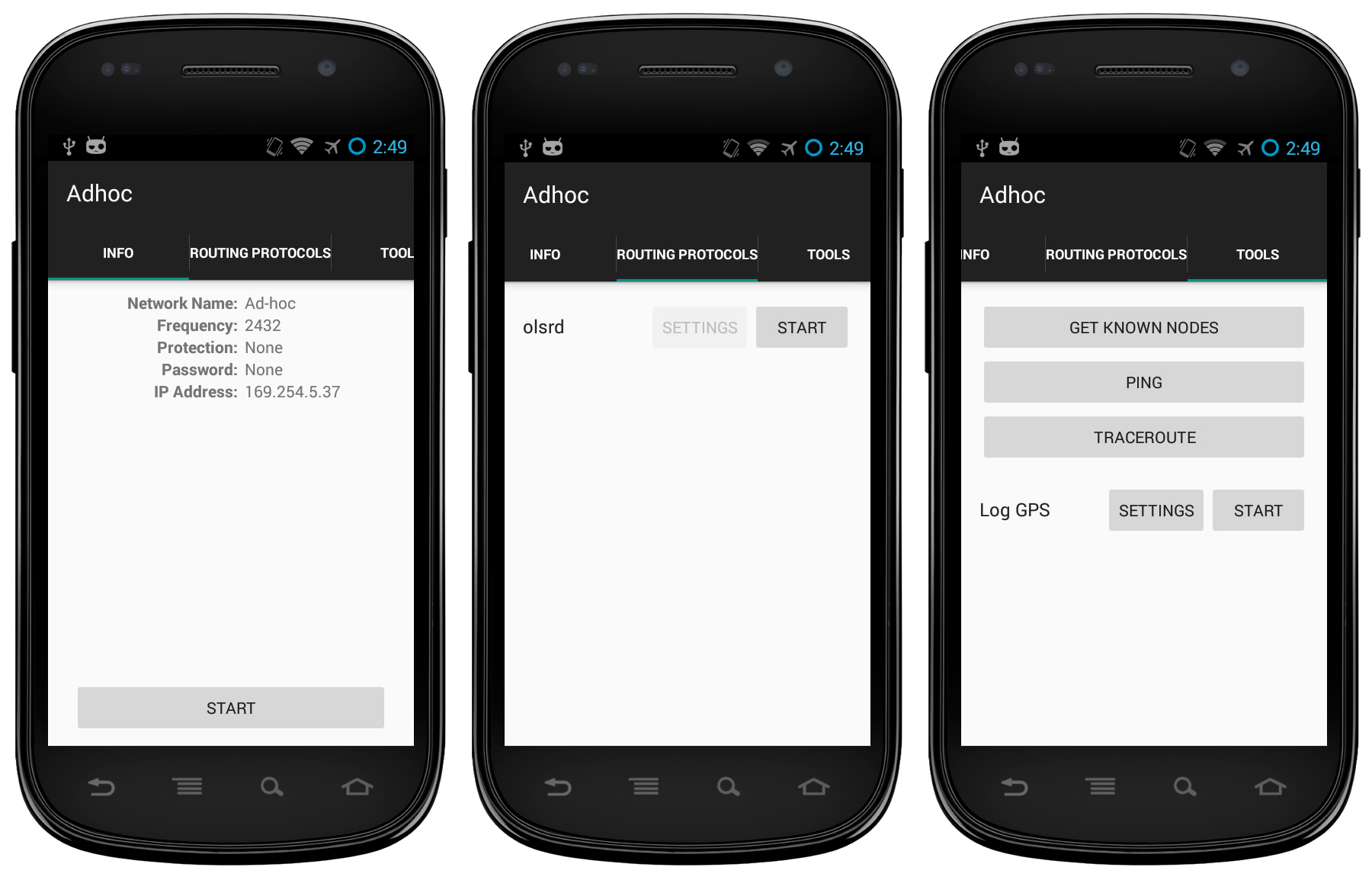}
    \caption{AdHocDroid application screen-shots}
    \label{fig:adHocDroid}
\end{figure}

AdHocDroid is an Android application that makes the necessary changes in the device to effortlessly create a \ac{MANET} in one step, as shown in \figref{adHocDroid}. The application sets up the \ac{IBSS} network, enabling ad-hoc mode on the wireless card, offers the possibility to choose the network name, and configures the IP address, network mask and gateway for the device. All parameters have default values, e.g.~the IP address is chosen according to~\cite{RFC3927}. 
The application also allows an easy way to import and run different routing protocols, and using tools to  monitor and evaluate the state of the network. 

We where able to successfully use our application enabling multi-hop connectivity on the Gigabyte Gsmart G1305 with Android~2.3 (CyanogenMod~7), and Samsung Nexus S with Android~4.3 (CyanogenMod~10.2.1) smartphones, and on the Samsung Galaxy Tab 10.1 tablet with Android~3.2. We also tested it on LG Nexus 4, LG Nexus 5 and Motorola Moto~G~(2013), but due to driver or chipset issues the Ad-Hoc mode did not function correctly. We have also tested the smartphone ad-hoc network with a 1st~responder monitoring application with success~\cite{Aguiar2014}.

\subsection{Architecture}
\begin{figure}[!ht]
	\centering
    \includegraphics[width=.95\linewidth]{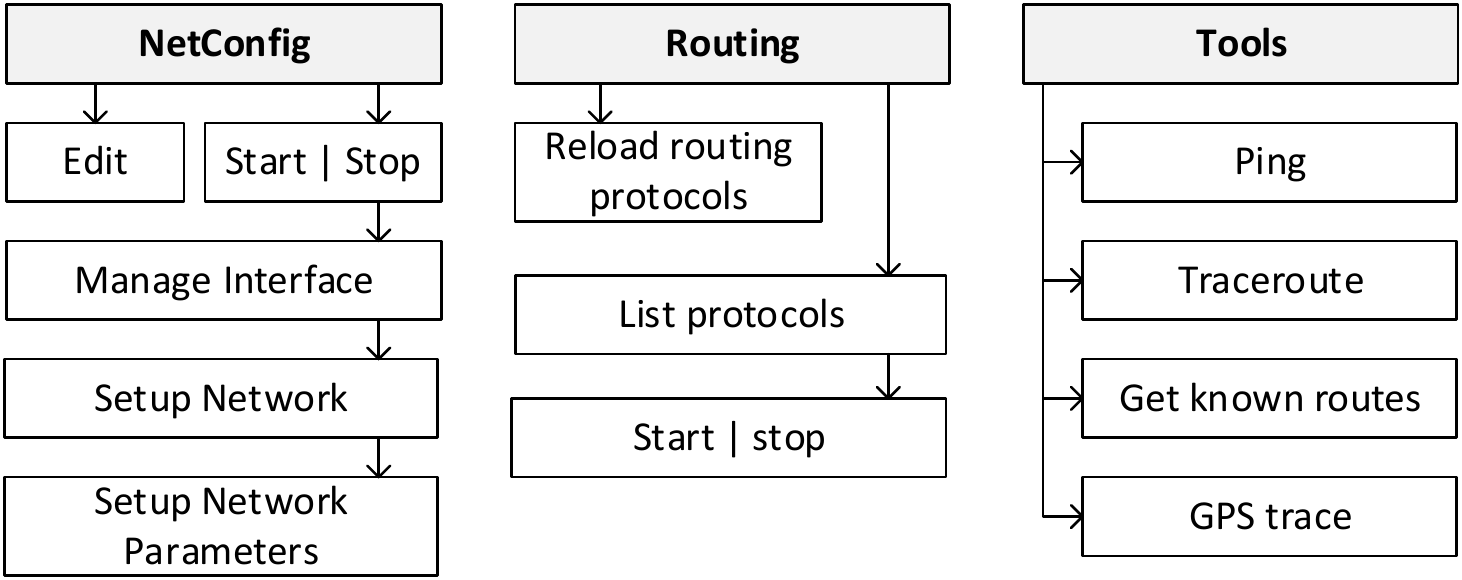}
    \caption{AdHocDroid application architecture}
    \label{fig:app_architecture}
\end{figure}

The application follows a modular architecture as shown in \figref{app_architecture}. There are three clearly defined modules that map to the application seen in \figref{adHocDroid} (left-to-right), the network configuration (NetConfig), the routing protocols (Routing) and tools (Tools).

For the network configuration all parameters can be adjusted. The start/stop button executes the network stack setup procedure. 
First it turns off the network interface via the Android \ac{API}, which avoids any other application (third party or even the system settings) making changes to the configurations of the networks. 
It then alters the text file that stores all known networks to add or remove the \ac{IBSS} network (the \texttt{wpa\_supplicant.conf}).
It finishes turning on the network interface in order to load the new network, and apply to it the network parameters (IP address, network mask, gateway).
This last step issues commands on the command line.

For multi-hop connectivity, we bundle the application with an \ac{OLSR}\footnote{See \url{http://www.olsr.org/}} routing daemon, but it is easy to import, start and stop other routing protocols. This can be done without recompiling or changing AdHocDroid, by creating a zip file with bash scripts for starting and stopping the routing protocol, and the executable binaries with the implementation of the protocol cross-compiled to the architecture of the device. We provide more detailed documentation on how to add and use the pre-compiled \ac{OLSR} build that runs on Android devices as an example in \url{{https://github.com/eSoares/Routing-Protocol-package-to-Android-Ad-hoc-framework}}. 
Using this routing protocol, we verified that it is possible for terminals running other \acp{OS} to join the created \ac{MANET}. This was tested with a laptop running a Linux distribution (Ubuntu) with \ac{OLSR} and two smartphones compatible with AdHocDroid. 

AdHocDroid has additional tools that we have found useful when carrying out experiments in the field, like providing information on the routing table, and execute ping or traceroute commands. It is also possible to log the smartphone \acs{GPS} coordinates, e.g. to map connectivity using geographic positions.



\subsection{Lessons Learned}
During development and testing, we came across problems with the Android \acp{API} and the diversity of devices, which we report in this section. 

The Android \ac{API} to configure network information, namely IP address, gateway and network mask, was deprecated in Android HoneyComb (3.0). To surpass this problem we used reflection through undocumented and internal \acp{API}, and wrapped this in a library, which we published in \url{https://github.com/eSoares/Android-IP-Manager}.

The Android \ac{API} does not enable the creation of an ad-hoc network. Thus, the library directly edits, as described, the  system configuration file, usually \texttt{wpa\_supplicant.conf} in Linux distributions, located in \texttt{/data/misc/wifi/} in Android. However, we found that it has other names in some devices. For example, in Samsung Galaxy Tab 10.1 it was named \texttt{bcm\_supp.conf} and was located in \texttt{/data/wifi/}.

Sometimes, after adding a new \ac{IBSS} network, the phone would prefer a previously saved 802.11 network instead of connecting to the new network. To fix this, the library edits \texttt{wpa\_supplicant.conf} to show only \ac{IBSS} networks to the \ac{OS} when we want to connect to ad-hoc networks. 

Even after the network was completely configured and set up, some devices (LG Nexus 5 and Motorola Moto G (2013)) were unable to connect to the \ac{MANET}. We assume that the WiFi chipset drivers did not implement this mode, since the Android \ac{OS} did not present any limitation and 
messages of issues in the driver where present in the Android WiFi state machine. 
%

Some other devices (LG Nexus 4) are able to connect to the network in \ac{IBSS} mode, but we found non-compliant behaviour. The first device to connect would act as an \ac{AP}, and other devices would from then on use this \ac{AP} to route traffic. If the first device (acting as \ac{AP}) left or went out of range, the rest of devices in the network where unable to communicate. 

In summary, \ac{IBSS} mode in Android devices is problematic mainly due to drivers or chipsets not implementing the required functionality. Nevertheless, we tested and were/are able to run AdHocDroid repeatedly and consistently on Gigabyte Gsmart G1305, Samsung Nexus S and Samsung Galaxy Tab~10.1. 



\section{Is It Really a Wireless MANET?}
\label{sec:manetcompare}
Given that some applications claim to provide mobile ad-hoc networking, we set off to think about what defines a \ac{MANET}. In our view, that definition requires answering all the following questions with yes:
\begin{itemize}
  \item is communication possible without connectivity to the Internet? (\textbf{No Internet Needed})
  \item is multi-hop communication possible? (\textbf{Multi-hop})
\item can any application take advantage of the provided connectivity through a regular socket \ac{API}, thus not requiring adaptation/re-writing? (\textbf{Any App})
%
\item can work without needing additional wireless technology to provide communication, e.g.: using IEEE 802.11 needs also \ac{BT}? (\textbf{No other Wireless})
\item can we use off-the-shelf \acp{OS} to communicate with the \ac{MANET}? E.g.: if development is on Android can we communicate with a PC running another \ac{OS}? (\textbf{Other Systems})
\end{itemize}

We analysed the technologies addressed in section~\ref{sec:related-work} according to this definition, and summarise the results in \tabref{comparison}. 

The main problem that almost every proposal faces is the support in different systems. This in some cases may be a matter of adoption (802.11s and AdHocDroid) while in others it involves "heavier" development from the proponents themselves. Some points are critical for a \ac{MANET} testbed namely supporting multi-hop and providing the regular socket interface for applications. In our view, this makes Open Garden, Serval and WiFi Direct definitively not fit the "wireless \ac{MANET}" name.

\begin{table}[htb]
\centering
\caption{\ac{MANET} network solutions check-list. \textit{yes} is denoted by {\cmark}, \textit{no} by \xmark{} and {\pmark} indicates \textit{partially} or \textit{with some adaptations}}
\scriptsize
\begin{tabular}{ |>{\columncolor{gray-cell}\centering}m{4.3em} |>{\centering}m{4.0em} |>{\centering}m{3.0em} |>{\centering}m{2.1em} |>{\centering}m{4.2em} |>{\centering}m{4.0em}|}
  \hline
  \textbf{Proposal}  & \textbf{No Internet Needed} & \textbf{Multi-hop} & \textbf{Any App} & \textbf{No other Wireless} & \textbf{Other Systems} 
  \tabularnewline
  \hline

  802.11s (native) & \cmarkcell & \cmarkcell & \cmarkcell & \cmarkcell & \pmarkcell \tabularnewline 
\hline

Open Garden & \cmarkcell & \pmarkcell & \xmarkcell & \xmarkcell  & \xmarkcell \tabularnewline 
\hline
  
 Serval & \cmarkcell & \cmarkcell & \xmarkcell  & \cmarkcell & \xmarkcell \tabularnewline 
\hline
  
  
 WiFi Direct & \cmarkcell & \xmarkcell & \xmarkcell & \cmarkcell & \cmarkcell \tabularnewline 
\hline

AdHoc-Droid & \cmarkcell & \cmarkcell & \cmarkcell & \cmarkcell & \pmarkcell \tabularnewline 
\hline
  \end{tabular}

\normalsize
\label{tab:comparison}
\end{table}

As is summarized in \tabref{comparison}, only AdHocDroid and 802.11s truly provide all the features for a \ac{MANET}. As we mentioned our aim is to advance the state-of-the-art in \ac{MANET} experimentation by providing a framework to easily test \ac{MANET} network protocols. In the future, 802.11s, when it becomes adopted in chipsets and their drivers, will provide an easy establishment of a multi-hop \ac{MANET}. However, as multi-hop is provided at driver level, it can be more difficult to test routing protocols than with our solution.

\section{Experimental Results}
\label{sec:performance-evaluation}

\begin{figure}[!ht]
	\centering
	\includegraphics[width=.95\linewidth, page=4]{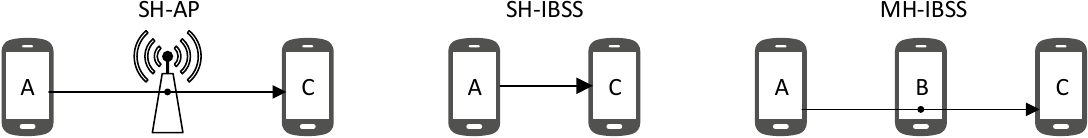}
    \caption{Test scenarios}
    \label{fig:scenarios}
\end{figure}

We performed a series of experiments to test AdHocDroid both from an end-user perspective, using standard applications (multiplayer games) from the Play Store to verify that they work unchanged in a \ac{MANET}, and from a technical perspective, measuring throughput, delay and battery consumption and comparing them with values obtained in a standard infrastructured scenario with an \ac{AP}.
The experiments were performed using three Samsung Nexus~S smartphones running Android~4.3 with CyanogenMod 10.2.1, in single (SH) and multihop (MH) configurations, as shown in Figure~\ref{fig:scenarios}. 

\subsection{Standard Applications}

From an end-user perspective, we downloaded from the Play Store two different multiplayer games for the local network, \emph{2048 BATTLE - multiplayer game}\footnote{\url{https://play.google.com/store/apps/details?id=com.jtataming.BATTLE2048}} and \emph{Spaceteam}\footnote{\url{https://play.google.com/store/apps/details?id=com.sleepingbeastgames.spaceteam}}, and tested them in single and multihop \ac{IBSS}. 
\emph{2048 BATTLE} worked straightforward. The application asks for the last byte of the IP address of the opponent (apparently assuming a class C network), and connecting the two users works even in the multihop scenario.  Changing between the single and multihop scenarios during the game did not disrupt it in any perceivable way.

In \emph{Spaceteam}, connecting the users worked as expected in single hop, but not in multihop.  This problem occurs because the game uses multicast DNS to find the opponent in the local network, but multicast traffic is not forwarded by the router (node~B).  After the discovery phase, the game worked normally, even if we changed from single to multihop and vice-versa during the game.  

\subsection{Performance Evaluation}

We carried out experiments to validate the ad-hoc multihop functionality measuring throughput and delay, and then quantify the impact of AdHocDroid on battery discharge, an important metric for user acceptance of MANET based applications.

We measured the throughput of the different nodes in the scenarios shown in Figure~\ref{fig:scenarios}. We generated traffic from node~A to node~C using iPerf (version~2.0.5)\footnote{\url{https://iperf.fr/}}, sending \ac{UDP} packets at the maximum 802.11a/g PHY rate (54~Mbits/s), thus overloading the channel and collecting throughput once per second. As the experiments were done while measuring battery discharge, they were done until one of the nodes had the battery depleted. We executed six of these batches.

\begin{figure}[!ht]
	\centering
    \includegraphics[width=.8\linewidth]{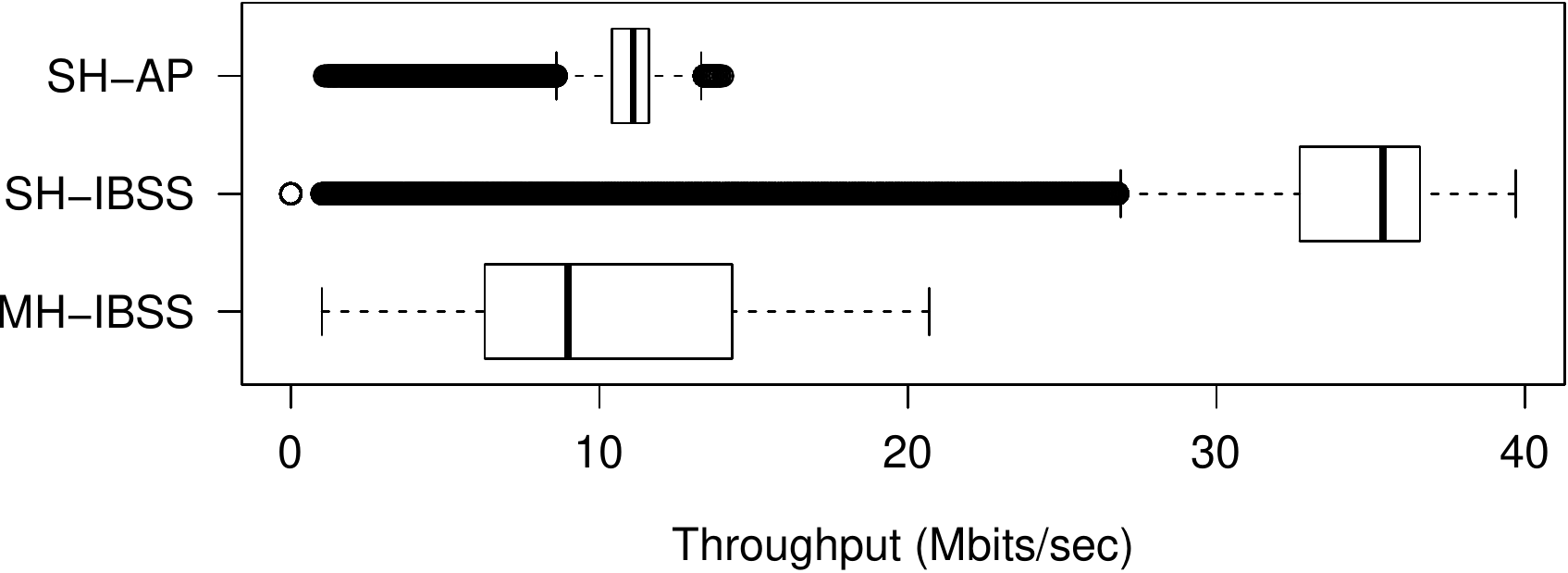}
    \caption{Throughput for infrastructure vs. \ac{IBSS} single hop vs. \ac{IBSS} multi-hop scenarios}
    \label{fig:throughput}
\end{figure}

The results are shown in Figure~\ref{fig:throughput}.  As expected, throughput is comparable in the multihop \ac{IBSS} and the infrastructured cases, and larger (more than double) in the single hop \ac{IBSS} case. This is due to the absence of forwarding at an intermediate node (\ac{AP} or router), implying lower medium contention.

\begin{figure}[!h]
	\centering
    \includegraphics[width=0.8\linewidth]{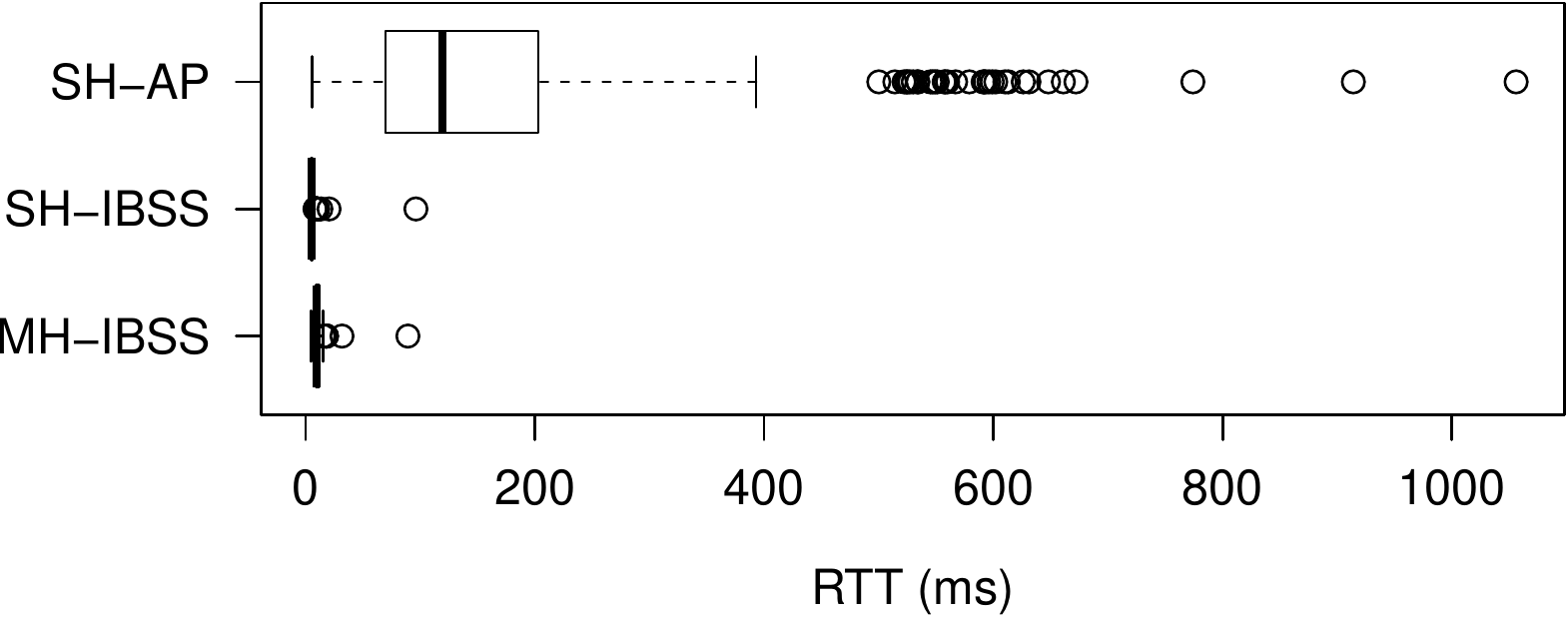}
%
    \caption{Ping RTT for infrastructure vs. \ac{IBSS} single hop vs. \ac{IBSS} multi-hop scenarios between node A to C}
    \label{fig:pings}
\end{figure}

To evaluate the end-to-end delay, we measured the \ac{RTT} using \emph{ping}, sending one request per second for 30 seconds.  We ran six series with a one minute interval between series.  Figure~\ref{fig:pings} shows the results.  In \ac{IBSS} mode, the \ac{RTT} in multihop is about twice as large as in single hop, as expected.  Through an \ac{AP}, the \ac{RTT} was much larger, ranging up to 400~ms (with the exception of some larger valued outliers).  This was consistent across several models of \ac{AP} (including a smartphone configured for wireless tethering but disconnected from the Internet).  This large difference is probably due to the absence of power saving in \ac{IBSS} mode, which is consistent with the difference in battery consumption without network traffic that we report below. Note that this does not apply to the tests with traffic, because nodes with permanent backlog do not enter
the power save state. These and the previous results validate the correct functioning of the smartphone MANET.

\begin{figure}[!ht]
	\centering
	\includegraphics[width=.8\linewidth]{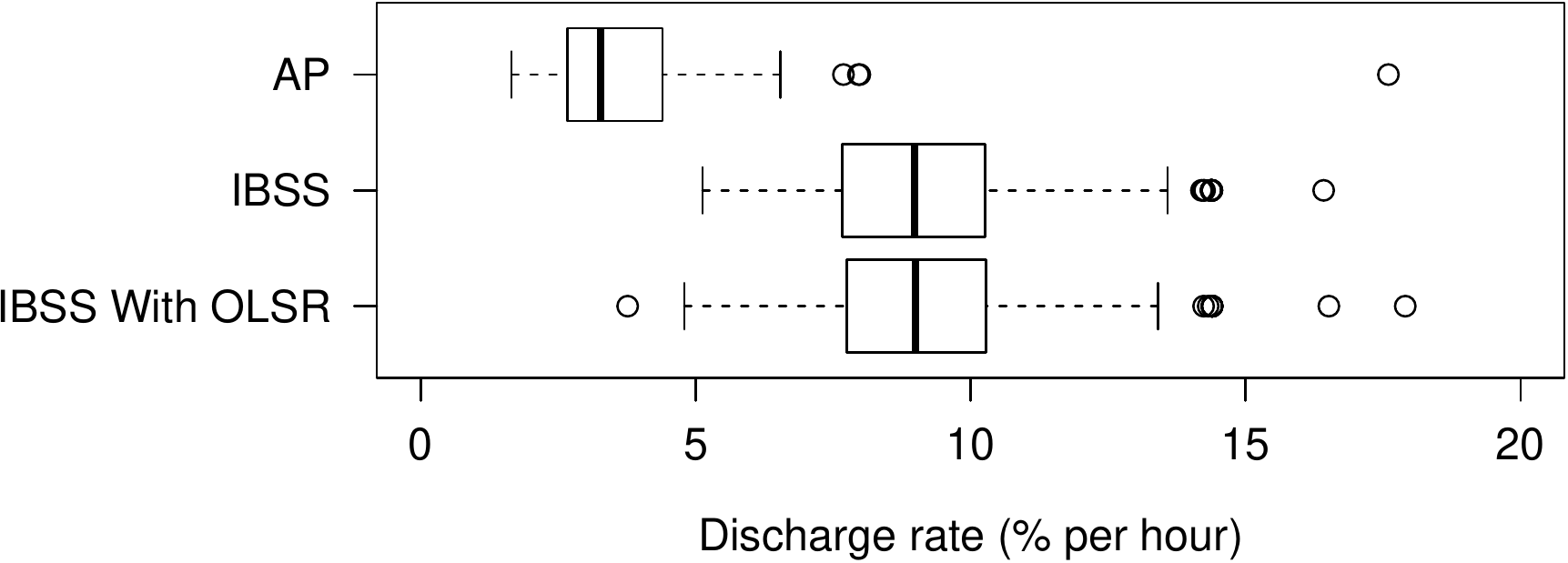}
	\caption{Discharge rate for infrastructure vs. \ac{IBSS} without OLSR vs. \ac{IBSS} with OLSR}
	\label{fig:batteryNoTraffic}
\end{figure}

Finally, we evaluated the impact on a smartphone's battery discharge rate of being used as \ac{MANET} node. We started by measuring the battery impact of simply setting the devices to \ac{IBSS} mode, without network traffic. The results are shown in Figure~\ref{fig:batteryNoTraffic}. Compared to an infrastructure scenario through an \ac{AP}, there is a significant increase in battery consumption, however the \ac{OLSR} daemon with the standard configurations does not further impact this consumption (note that this is a stationary scenario with only three nodes).

\begin{figure}[!ht]
	\centering
    \includegraphics[width=.8\linewidth]{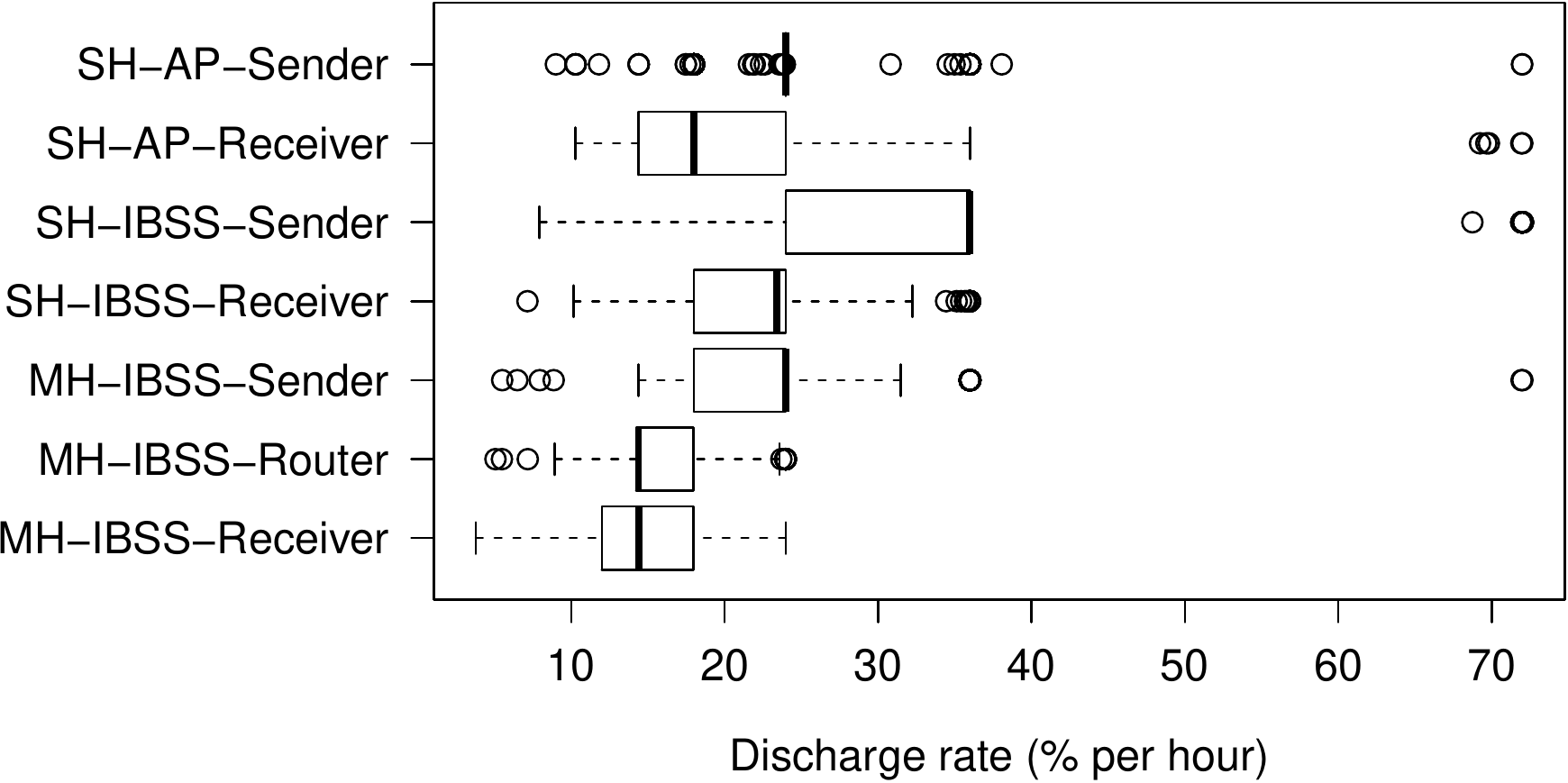}
    \caption{Discharge rate for infrastructure vs. \ac{IBSS} single hop vs. \ac{IBSS} multi-hop scenarios}
    \label{fig:battery}
\end{figure}

We also measured the discharge rate of the different nodes in the scenarios shown in Figure~\ref{fig:scenarios}. 
Because the discharge rate is a function of the amount of traffic being sent\slash received, overloading the channel, as iPerf does, ensures that we are observing worst-case battery discharge rates. We used the same setting as in the throughput experiments, as mentioned. In these, we additionally collected the time interval for each percent point drop in battery until the first node has the battery depleted. We plot the distribution of the discharge rate between consecutive points in Figure~\ref{fig:battery}.

Battery consumption in multihop \ac{IBSS} is comparable to the infrastructured case for both sender and receiver. In single hop \ac{IBSS}, the consumption is higher since the throughput is also higher due to the lower medium contention.
  Consumption at the sender is higher than at the receiver because it is trying to send at a higher rate than is actually possible. The router in the multihop \ac{IBSS} scenario discharges at a similar rate as the receiver, confirming that the discharge rate is determined by the time that the network interface card is busy. We double checked this by sending at rate lower than saturation (6~Mbps) and observing similar battery discharge rates in all nodes. 
\needspace{2\baselineskip}
\section{Conclusions}\label{sec:conclusions}
Motivated by a wide range of application scenarios for MANETs of smartphones proposed in the literature, we survey the existing solutions to actually enable IP-level ad-hoc networking on Android smartphones. 
We describe a software to enable ad-hoc networking on Android devices and shared learned lessons, so that anyone in the community can easily build a \ac{MANET} test-bed with multihop capability that applications can access through the sockets \ac{API}. 
We thus expect to contribute to moving the envisioned applications one step closer to reality.

As an example of the want of developers and others for ad-hoc support, we have that "Support Wi-Fi ad-hoc networking" was \href{https://code.google.com/p/android/issues/detail?id=82}{trouble ticket 82}\footnote{\url{https://code.google.com/p/android/issues/detail?id=82}} in the Android open source project, starred by 6300 people until it was classified as obsolete and closed in April 1st 2015.  For comparison, the most starred ticket related to networking had been starred 9751 times and the second most starred ticket 3565 times, as of Oct 2015.



\section*{Acknowledgements}
Research funded by Project VR2Market - Ref. CMUP-ERI/FIA/0031/2013.

\bibliographystyle{IEEEtran}
\bibliography{IEEEabrv,biblio}

\end{document}